\documentclass{iauc}
\usepackage{graphicx}

\title[Migration of small bodies and dust] 
{Migration of small bodies and dust to the terrestrial planets}

\author[S.I. Ipatov and J.C. Mather]   
{Sergei I. Ipatov$^{1,2}$ and John C. Mather$^3$}%

\affiliation{$^1$Catholic University of America, USA 
\break 
email: siipatov@hotmail.com \\[\affilskip]
$^2$Institute of Applied Mathematics, Moscow, Russia \\[\affilskip]
$^3$LASP, NASA/Goddard Space Flight Center, Greenbelt, USA
\break 
email: John.C.Mather@nasa.gov}

\pubyear{2004}
\volume{197} 
\pagerange{ -- }
\date{?? and in revised form ??}
\jname{Dynamics of Populations of Planetary Systemsd}
\editors{Z. Knezevic and A. Milani, eds.}
\begin{document}

\maketitle

\begin{abstract}
We integrated the orbital evolution of 30,000 Jupiter-family comets, 1300 resonant asteroids, and 7000 asteroidal, trans-Neptunian, and cometary dust particles.
For initial orbital elements of bodies close to those of Comets 2P, 
10P, 44P, and 113P, a few objects  got 
Earth-crossing orbits with semi-major axes $a$$<$2 AU and 
moved in such orbits 
for more than 1 Myr (up to tens or even hundreds of Myrs). 
Four objects (from 2P and 10P runs) even got inner-Earth 
orbits (with aphelion distance $Q$$<$0.983 AU) and 
Aten orbits for Myrs. 
Our results show that the trans-Neptunian belt can provide a
significant portion of near-Earth objects, or the number of trans-Neptunian objects migrating inside the solar system can be smaller than it 
was earlier considered, or most of 1-km former trans-Neptunian objects that 
had got near-Earth object orbits for millions of years disintegrated into mini-comets and dust during a smaller part of their dynamical lifetimes.
The probability of a collision of an asteroidal or cometary particle during its lifetime with the Earth
was maximum at diameter $d$$\sim$100 $\mu$m. At $d$$<$10 $\mu$m such probability for trans-Neptunian particles was less than that for asteroidal particles by less than an order of magnitude, so the fraction of  trans-Neptunian particles with such diameter near Earth can be considerable.


\end{abstract}

\firstsection 
\section{Introduction}

Celestial bodies and dust particles migrate to near-Earth space from different regions of the solar system (from the main asteroid and Edgeworth-Kuiper belts, the Oort and Hills clouds, etc.). Some scientists (Farinella et al. 1993; Bottke et al. 2002) considered that most near-Earth objects (NEOs) are asteroidal fragments, others (Wetherill 1988) supposed that half of NEOs are former short-period comets. We studied migration of small bodies and dust particles based on extensive integrations, considering more Jupiter-crossing objects than before. The interval of considered diameters of dust particles was wider than in previous publications. Our reviews of papers on these problems can be found in (Ipatov 2001, Ipatov \& Mather 2004a-b, Ipatov et al. 2004a).

\section{Migration of Jupiter-family comets to the Earth}

As migration of trans-Neptunian objects (TNOs) to Jupiter's orbit was investigated by several authors (e.g., Levison \& Duncan, 1997), Ipatov \& Mather (2003, 2004a-b) integrated the orbital evolution of ~30,000 Jupiter-family comets (JFCs) under the gravitational influence of planets, and based on the results of these integrations they investigated the migration of TNOs to the near-Earth space. Also the orbital evolution of 1300 resonant main-belt asteroids was studied. We omitted the influence of Mercury (except for Comet 2P/Encke) and Pluto and used the Bulirsch-Stoer and symplectic methods (BULSTO and RMVS3 codes) from the integration package of Levison and Duncan (1994). In the first series of runs (denoted as $n1$), we calculated the evolution of 3100 Jupiter-crossing objects (JCOs) moving in initial orbits close to those of 20 real comets with period 5$<$$P_a$$<$9 yr, and in the second series of runs (denoted as $n2$) we considered 13,500 JCOs moving in initial orbits close to those of 10 real comets with period 5$<$$P_a$$<$15 yr. In other series of runs, initial orbits were close to those of a single comet (2P, 9P, 10P, 22P, 28P, 39P, or 44P). We investigated the orbital evolution during the dynamical lifetimes of objects (at least until all the objects reached perihelion distance $q$$>$6 AU).

In our runs, planets were considered as material points, so literal collisions did not occur. 
However, based on the orbital elements sampled with a 500 yr step, we calculated the mean probability $P$ of collisions during lifetime of a JCO. 
The results can differ considerably depending on the initial orbits of comets. The values of $P$ 
for the Earth were about (1-4)$\cdot$$10^{-6}$ for Comets 9P, 22P, 28P, and 39P. They were (6-10)$\cdot$$10^{-6}$ for Comet 10P and exceeded $10^{-4}$ for Comet 2P.
The ratio of the mean probability of a JCO with 
semi-major axis $a$$>$1 AU with a planet to the mass of the planet was greater for Mars by a factor of several than that for Earth and Venus.
Using $P$=$4\cdot10^{-6}$ (this value is smaller than the mean value of $P$
obtained in our runs) and assuming that 
the total mass of planetesimals that ever crossed Jupiter's orbit is  about 100$m_E$ (Ipatov 1987, 1993), 
where $m_E$ is the mass of the Earth,  we obtained that the total mass of water 
delivered from the feeding zone of the giant planets to the Earth could be about 
the mass of water in Earth's oceans. Ancient Venus and Mars could have large oceans.

For series $n1$ with RMVS3, the probability of a collision with the Earth for one object with 
initial orbit close to that of Comet 44P was 88.3\% of the total probability 
for 1200 objects from this series, and the total probability for 1198 objects was only 4\%. 
The probabilities of collisions of one object 
with Earth and Venus could be greater than for thousands other objects combined.


A few JFCs got Earth-crossing orbits with 
aphelion distance $Q$$<$4.2 AU (and sometimes with $a$$<$2 AU)
and moved in such orbits for more than 1 Myr (up to tens or even hundreds of Myrs). The probability of a collision of one of such objects, which move for millions of years inside Jupiter's orbit, with a terrestrial planet can be greater than analogous total probability for thousands other objects. 
With BULSTO, six and nine objects, respectively from 10P and 2P series, 
moved into Apollo orbits with $a$$<$2 AU  for at least 0.5 Myr each, and five of them remained in such orbits for more than 5 Myr each. 
Only two objects in series $n2$ got Apollo orbits with $a$$<$2 AU during more than 1 Myr. 

Four considered former JFCs even got inner-Earth object (IEO) orbits (with $Q$$<$0.983 AU) or Aten orbits for Myrs. Note that Ipatov (1995) obtained migration of JCOs into IEOs using the method of spheres to consider the gravitational influence of planets. Now let’s return to our recent results of integration. One former JCO got Aten orbits ($a$$<$1 AU, $Q$$>$0.983 AU) during $>$3 Myr, but a probability of its collision with the Earth was greater than that for $10^4$ other considered former JCOs. This object also moved during about 10 Myr (before its collision with Venus) in IEO orbits, and during this time interval a probability of its collision with Venus was even greater than the total probability of collisions of $10^4$ considered JCOs with Earth. 

After 40 Myr one considered object with initial orbit close to that of Comet 88P got  $Q$$<$3.5 AU, and it moved in orbits with  $a$=2.60-2.61 AU,  1.7$<$$q$$<$2.2 AU, 3.1$<$$Q$$<$3.5 AU, eccentricity $e$=0.2-0.3, and inclination $i$=5-10$^\circ$ for 650 Myr. Another object (with initial orbit close to that of Comet 94P) moved in orbits with $a$=1.95-2.1 AU, $q$$>$1.4 AU, $Q$$<$2.6 AU, $e$=0.2-0.3, and $i$=9-33$^\circ$ for 8 Myr (and it had $Q$$<$3 AU for 100 Myr). So JFCs can rarely get typical asteroid orbits and move in them for Myrs. In our opinion, it can be possible that Comet 133P (Elst-Pizarro) moving in a typical asteroidal orbit was earlier a JFC and it circulated its orbit also due to non-gravitational forces. 
JFCs got typical asteroidal orbits less often than NEO orbits. 

Results of the orbital evolution of JCOs show that many Earth-crossing objects can move in highly eccentrical ($e$$>$0.6) orbits and, probably, most of 1-km objects in such orbits have not yet been discovered. The obtained results show that during the accumulation of the giant planets the total mass of icy bodies delivered to the Earth could be about the mass of water in Earth's oceans.

The results obtained by the Bulirsh-Stoer method (BULSTO code) with an integration 
step error $\epsilon_r$ $\sim$10$^{-9}$-10$^{-8}$, with 10$^{-12}$, and by a 
symplectic method (RMVS3 code) with an integration step $d_s$$\le$10 days were 
usually similar, and the difference at these three series was about the difference 
at small variation of $\epsilon_r$ or $d_s$. In the case of close encounters with 
the Sun, the values of collision probability $P_S$ with the Sun obtained by BULSTO 
and RMVS3 and at different $\epsilon_r$ or $d_s$ could be different, but all other 
results were usually similar. For Comet 2P, Comet 96P, and the 3:1 resonance with Jupiter, 
close encounters with the Sun took place, and with a symplectic method we got greater mean probabilities of 
collisions with the Sun than with BULSTO. 
The fraction of asteroids migrated from the 3:1 resonance with Jupiter 
that collided with the Earth was greater by a factor of several than that 
for the 5:2 resonance.

\section {Trans-Neptunian objects in near-Earth object orbits}

Using our results of the orbital evolution of JCOs and the results of 
migration of TNOs obtained by Duncan et al. (1995) and 
considering the total of $5\cdot10^9$ 1-km TNOs with  30$<$$a$$<$50 AU, 
Ipatov \& Mather (2003, 2004a-b) estimated the number of 1-km former TNOs in NEO orbits. 
Results of our runs testify in favor of at least one of these conclusions: 1) the portion of 1-km former TNOs among NEOs can exceed several tens of percents, 2) the number of TNOs migrating inside the solar system could be smaller by a factor of several than it was earlier considered, 3) most of 1-km former TNOs that had got NEO orbits disintegrated into mini-comets and dust during a smaller part of their dynamical lifetimes if these lifetimes are not small. 

Our runs showed that if one observes former JFCs in NEO orbits, then most of them could have already moved in such orbits for millions (or at least hundreds of thousands) of years (if they didn't disintegrate). Some former comets that have moved in typical NEO orbits for millions or even hundreds of millions of years, and might have had multiple close encounters with the Sun, could have lost their typically dark surface material, thus brightening their low albedo and assuming the aspect typical of an asteroid (for most observed NEOs, the albedo is greater than that for comets). The number of extinct comets can exceed the number of active comets by several orders of magnitude if most of comets do not disintegrate into 
mini-comets and dust during their dynamical lifetimes.

\section{Migration of dust particles to the terrestrial planets}

Using the Bulirsh-Stoer method of integration, Ipatov et al. (2004a) integrated the orbital evolution of 7000 asteroidal, trans-Neptunian, and cometary dust particles, under the gravitational influence of planets, the Poynting-Robertson drag, radiation pressure, and solar wind drag. The values of the ratio $\beta$ between the radiation pressure force and the gravitational force were varied from 0.0004 to 0.4. For silicates, such values correspond to particle diameters $d$ between 1000 and 1 microns; for water ice, the diameters are greater by a factor of 3 than those for silicates. After the above publication, we have made more calculations for small values of $\beta$ and more analysis of the runs. Considered asteroidal and trans-Neptunian (kuiperoidal) dust particles started with zero relative velocity from the numbered main-belt asteroids and known trans-Neptunian objects, respectively. We also studied particles started from Comet Encke.
Several hundred of particles were usually considered at each $\beta$. 
The simulations continued until all of the particles either collided with the Sun or 
reached 2000 AU from the Sun. In our runs, planets were considered as material points, but using orbital elements 
obtained with a step $dt$ of $\le$20 yr, we calculated the mean probability $P$
of a collision of a particle with a planet during the lifetime of the particle. 
We also made similar runs without planets to investigate the role of planets in 
interplanetary dust migration.


The mean probabilities $P$ of collisions of asteroidal and cometary dust particles with Earth and Venus during lifetimes of particles were maximum at $\beta$$\sim$0.002-0.004 (i.e., at $d$$\sim$100-200 microns for silicate particles). At $\beta$$\ge$0.01 the values of $P$ and the mean times $T$ spent by particles in orbits with $q$$<$1 AU  quickly decrease with an increase of $\beta$  (usually $P$ and $T$ are proportional to 1/$\beta$). For asteroidal dust particles at $\beta$$\sim$0.0004-0.001 the values of $P$ and $T$  were smaller than those at $\beta$$\sim$0.002-0.004, though maximum times elapsed until collisions of particles with the Sun were greater for smaller $\beta$  (greater times were needed for larger particles to migrate to the orbits of the terrestrial planets). 
The probability of a 
collision of a migrating dust particle with the Earth for $\beta$$\le$0.01 is greater 
by a factor $>$200 than that for $\beta$=0.4. 
Cratering records in lunar material and on the panels of the Long Duration Exposure 
Facility showed that the mass distribution of dust particles encountering Earth 
peaks at $d$=200 $\mu$m.
The probability of collisions of cometary particles with the Earth is smaller than that for asteroidal particles at the same $\beta$, and this difference is greater for larger particles. 

Almost all asteroidal particles with $d$$\le$4 $\mu$m collided with the Sun. At $\beta$$\ge$0.02 and $\beta$$\le$0.001 some asteroidal particles migrated beyond Jupiter’s orbit. A few asteroidal particles can collide with the Sun after moving outside Jupiter's orbit for a long time. 


At $\beta$$\ge$0.05 (i.e., at $d$$<$10 $\mu$m) 
the probability of a collision of a trans-Neptunian particle with the Earth
and the mean time $T$ spent in orbits with $q$$<$1 AU
were usually less than those for a typical asteroidal particle by less than an order of magnitude (only at $\beta$=0.05 values of $T$ differed by a factor of 20). The total mass 
of the Edgeworth-Kuiper belt is at least two orders of magnitude greater than that 
of the main asteroid belt. So the fraction of trans-Neptunian particles with $d$$<$10 $\mu$m 
(runs with greater $d$ have not yet been finished)
among all particles near the orbit of the Earth can be considerable. 
At $\beta$$\sim$0.05-0.4, 
the fraction $P_S$ of trans-Neptunian particles collided with the Sun was 0.1-0.2, i.e., it was smaller by a factor of 5 than that of asteroidal particles for the same $\beta$.


Several plots of the distribution of migrating asteroidal particles in their orbital elements 
and the distribution of particles with their distance $R$ from the Sun and their height $h$ 
above the initial plane of the Earth's orbit were presented by Ipatov et al. (2004a). 
For $\beta$=0.01 the local maxima of 
the mean time $t_a$ (the total time divided by the number $N$ of particles) 
during which an asteroidal dust particle had a semi-major axis $a$ in an interval of fixed 
width  corresponding 
to the 6:7, 5:6, 3:4, and 2:3 resonances with the Earth are greater than the maximum at 2.4 AU. 
There are several other local maxima corresponding to the n:(n+1) resonances with Earth and 
Venus (e.g., the 7:8 and 4:5 resonances with Venus). The trapping of dust particles in the 
n:(n+1) resonances causes Earth's dust ring. The greater the $\beta$, 
the smaller the local maxima corresponding to these resonances. At $\beta$$\le$0.1 there are gaps with 
$a$ a little smaller than the semi-major axes of Venus and Earth that correspond to the 1:1 
resonance for each; the greater the $\beta$, the smaller the corresponding values of $a$. 
A small gap for Mars is seen only at $\beta$$\le$0.01. There are also gaps corresponding 
to the 3:1, 5:2, and 2:1 resonances with Jupiter. 

Spatial density $n_s$ of considered trans-Neptunian particles near ecliptic
at distance from the Sun $R$=1 AU was greater than at $R$$>$1 AU. 
At 0.1$\le$$\beta$$\le$0.4 and 2$<$$R$$<$45 AU 
(at $\beta$=0.05 for $11$$<$$R$$<$50 AU), $n_s$ varied with $R$ by less than a factor of 4.
For asteroidal and cometary particles, $n_s$ quickly decrease with an increase of $R$,
e.g., for $\beta$=0.2 at $R$=5 AU $n_s$ was smaller than at $R$=1 AU
by a factor of 70 and 50 for asteroidal and cometary particles, respectively. 

In contrast to the 
asteroidal dust particles, for cometary dust particles the values of a mean time in a planet-crossing orbit 
did not differ much between Venus, Earth, and Mars. 
Collision probabilities with Earth were greater by a factor of 10-20 than those 
with Mars and greater for particles starting at perihelion than aphelion. 
For the same values of $\beta$, the probability of cometary dust particles colliding 
with a terrestrial planet was several times smaller than for asteroidal dust particles, 
mainly due to the greater eccentricities and inclinations of the cometary particles. 
This difference is greater for larger particles. 
In almost all cases, inclinations $i$$<$$50^{\circ}$, and at $a$$>$10 AU the maximum $i$ was less for larger asteroidal particles. 


Based on the obtained distributions of migrating dust particles, 
Ipatov et al. (2004b) investigated how the solar spectrum is 
changed by scattering by the zodiacal cloud grains.  
As positions of particles were stored with a step of 10 yr, for each such 
stored position we calculated many ($>$10$^3$) different positions of a particle and the 
Earth considering that orbital elements do not vary during a period of 
revolution of the particle around the Sun. 
One of the 
obtained results is that the results of modeling are relatively 
insensitive to the scattering function considered.
In some models, the scattering 
function depended on a scattering angle $\theta$ in such a way: 
1/$\theta$ for $\theta$$<$c, 1+($\theta$-c)$^2$ for $\theta$$>$c, 
where c=2$\pi$/3 radian. In other runs, we added the same dependence 
on elongation $\epsilon$ (considered westward from the Sun). 
In the third runs, the scattering function didn't depend on 
these angles at all. 

The plots of the obtained spectrum are in general agreement with the 
observations made by Reynolds et al. (2004). 
They measured the profile of the scattered solar Mg I$\lambda$5184 absorption 
line in the zodiacal light.
Unlike results by Clarke et al. (1996), our modeled spectra do 
not exhibit strong asymmetry. 
As these authors, we obtained that minima in the plots of dependencies 
of the intensity of light on its wavelength near 5184 Angstrom are not so 
deep as those for the initial solar spectrum.
The details 
of plots depend on diameters, inclinations, and a source of particles.
For example, 
at inclination $i$=0 and $\epsilon$=90$^\circ$, for kuiperoidal and cometary 
particles the shift of the plot to the blue and its minimum were greater 
than those for asteroidal particles. 
Our preliminary models and comparison with observational data indicate that 
for more precise observations it will be possible to distinguish well 
the sources of the dust and impose constrains on the particle size. 

For asteroidal and cometary particles at $\beta$=0.2, about 65-80\% of brightness 
is due to particles at distance  from the Earth $r$$<$1 AU
(90-95\% for $r$$<$1.5 AU), and at most values of $\epsilon$ 
(not close to 180$^\circ$) 75-90\% of brightness is from particles with $R$$<$1.5 AU.
For trans-Neptunian particles, these fractions were smaller. 
At $\beta$=0.2 for $\epsilon$=270$^\circ$ about half of brightness was from particles
with $R$$>$4 AU and $r$$>$4 AU; for $\epsilon$=90$^\circ$  30-40 \% 
(depending on the considered scattering function) of brightness was from particles
with such values of $R$ and $r$.
Velocities of dust particles relative to the Earth that mainly 
contributed to brightness were different for different $\epsilon$. At $i$=0 
they were between -25 and 25 km/s.

\section{Conclusions}

Our results show that some former JFCs got Earth-crossing orbits with aphelion distance $Q$$<$4.2 AU and even inner-Earth object orbits (with $Q$$<$0.983 AU), Aten orbits, or typical asteroidal orbits for Myrs. The probability of a collision of one former JFC moving inside Jupiter's orbit for millions of years with a terrestrial planet can be greater than analogous total probability for thousands other JFCs. Results obtained by the Bulirsch-Stoer method were mainly similar (except for probabilities of close encounters with the Sun when they were high) to those obtained by a symplectic method. 
The total mass of icy bodies delivered to the Earth during the accumulation of the giant planets could be about the mass of water in Earth's oceans. 

We concluded that the trans-Neptunian belt can provide a significant portion of near-Earth objects, or the number of trans-Neptunian objects migrating inside solar system could be smaller than it was earlier considered, or most of 1-km former trans-Neptunian objects that had got near-Earth object orbits disintegrated into mini-comets and dust during a smaller part of their dynamical lifetimes if these lifetimes are not small. 

The probability of a collision of an asteroidal or cometary particle with the Earth
was maximum at diameter $d$$\sim$100 $\mu$m. At $d$$<$10 $\mu$m such probability for a trans-Neptunian particle was less than that for an asteroidal particle by less than an order of magnitude, and the fraction of  trans-Neptunian particles with such diameter near Earth can be considerable.
The peaks in the distribution of migrating asteroidal dust particles with semi-major axis corresponding to the n:(n+1) resonances with Earth and Venus and the gaps associated with the 1:1 resonances with these planets are more pronounced for larger particles. 

\begin{acknowledgments}
This work was supported by INTAS (00-240) and NASA (NAG5-12265).
\end{acknowledgments}

\end{document}